\documentclass[acmlarge]{acmart}
\usepackage{algorithm2e}

\AtBeginDocument{%
  \providecommand\BibTeX{{%
    \normalfont B\kern-0.5em{\scshape i\kern-0.25em b}\kern-0.8em\TeX}}}

\begin{document}

\title{Reputation System : Fair allocation of points to the editors in the collaborative community}

\author{Shubhendra Pal Singhal}
\authornotemark[1]
\email{106116088@nitt.edu}

\affiliation{%
  \institution{ \\ Department of Computer Science and Engineering\\National Institute of Technology, Tiruchirappalli\\}
}

\begin{abstract}
 In this paper, we are trying to determine a scheme for the fair allocation of points to the contributors of the collaborative community. The major problem of fair allocation of points among the contributors is that we have to analyze the improvement in the versions of an article. Let's say there is a contribution of major change in content which is relevant vs the contribution of adding a single comma. Every contributor cannot be given the same points in such a case. There are many ways which can be used like number of changes in a new version. That might seem relevant but it becomes irrelevant in terms of correct content contribution and other significant changes. There is no AI system too which can detect such a change and award the points accordingly. So, this problem of allocation of points to the contributors is presented by an algorithm with a theoretical proof. It relies on the interactive interaction of the users in the system which is trivial in case of big system design economies.

\end{abstract}
\keywords{Reputation System, Collaborative Community, Publishing, Article, Grading System}
\acmDOI{}
\renewcommand\footnotetextcopyrightpermission[1]{} 
\pagestyle{plain} 
\settopmatter{printacmref=false}
\setcopyright{none}
\renewcommand\footnotetextcopyrightpermission[1]{}
\pagestyle{plain}
\maketitle

\section{Introduction}
\subsection{System of Collaborative Community}
Collaborative community is an article publishing platform where contributors can publish, improve or suggest any article in their own space. The content of an article can be plain text or text with videos, images, and links. Users can create groups within the community, join group, add their content: mainly articles or collated articles. One user can join many different groups depending on his interest. Each group has a different role for a user - author, publisher. Author is the one who contributes to the community and publisher, in addition reviews the articles written by the user. A user  is allocated a role of publisher or author depending on his level and hours of contribution. The content submitted by the user for the review will be "visible" to the users of that community only. There can be versions of the same content indicating the improvement by any user or the implementation of the publisher's suggestion. If the publisher approves of the content, then the article is made public for any user of any community.\\
Groups and communities might need a private space for preparing an article. So the system for collaborative communities also provides a feature for a community to be a private where the article can be viewed by any user only after its publication is approved.\\
The system has a grading system but this grading varies depending upon the roles (author, publisher) and popularity of the article quantified by number of votes and views. Statistics involved in the fair allocation of the points to the community like views and votes is collected by the reputation system micro-service. It allows to present the  actual statistics about the active participation of users in the community.\\ 
\begin{figure}
 \includegraphics[scale = 0.8]{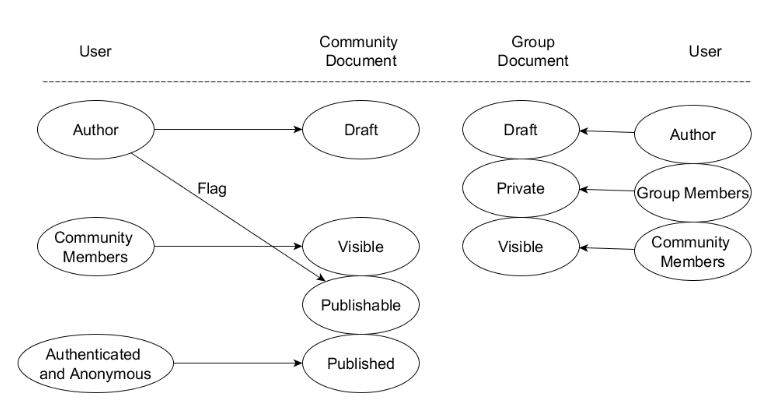}
 \caption{System of Collaborative Community}
\end{figure}
\subsection{Role Allocation and Dependency in Collaborative Community}
The super-admin is the creator for whole platform of collaborative system. Role of the super-admin is to approve a request of creation of community by a user and he can approve this request if and only if there is no other community existing with that same concept. The community created will be tagged with all those names which can belong to the same community, thus avoiding the confusion of how community creation can be wisely judged by just a search.\newline.
The authenticated users who will be interested in a community (by seeing their profiles) can join and draft/edit the content. One user can always be in multiple groups and then improve existing articles or content. The communities, thus formed, will have the roles of author ,publisher and community-admin. Those who draft the article or any content have the option of either privatizing their groups or they can make it public so that everyone can collaborate. If the group is made public then group-admin has the power to just remove them. But if it is private then the group-admin will be the first person in the group and has the control of granting permissions of entering and removing the group. The people who feel that they can contribute can join the group. Group in turn has just two roles open. First is the group-admin,then users who contribute to make the article reach , publishable state and last is the author.
As a user can be in more than one community, two types of reputations are maintained :\newline
1.) Community Reputation : The reputation of user in a particular community which has been obtained based on the grading system given under section 1.2 for community.\newline
2.) System Reputation : Article in Stack overflow corresponds to one community in our system and user participation might differ for different communities,so the allocation of points should be different.
\newline
The reputation that is the cumulative of all the community reputation that user is a part of.\\
\begin{figure}
  \includegraphics[scale = 0.4]{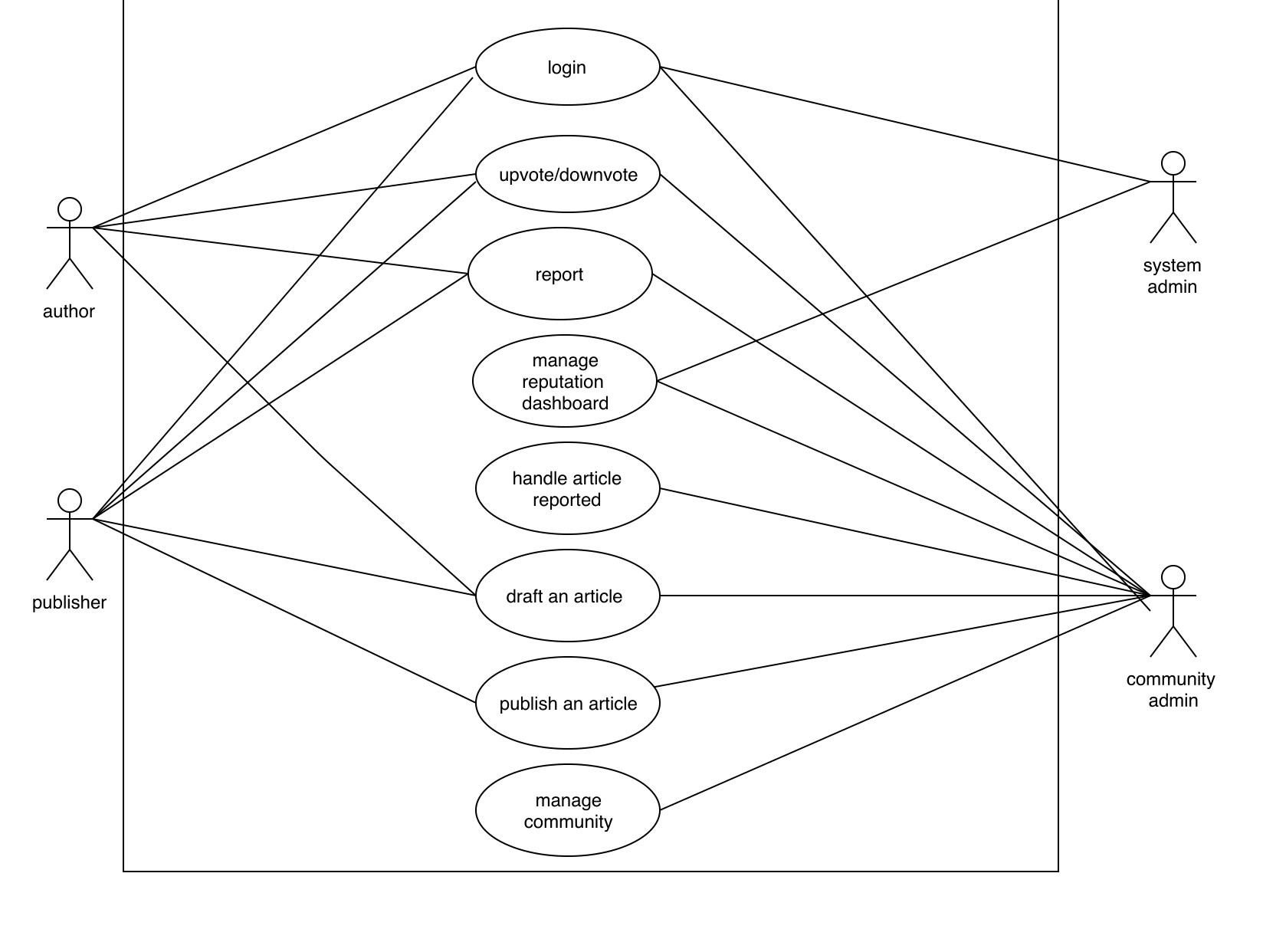}
  \caption{Roles and their respective responsibilities in the system }
\end{figure} 

\section{Research Model}
\subsection{Grading System}
The upvote in an article is a measure of how good is the version of the article and downvote is a measure of how can a version improved or negatively rated[1].
The point allocation system has been developed considering various factors including the chances of malpractices and certainty of cracking the system.
\newline
The point system is as follows:
\newline
Creating community (approved by the superadmin):+25\newline
Article published : +5 to the user of last edit and to others(explained in section 2.4), +5 to publisher.\newline
Article Reportedwrong(Approved) : -5 to the user of that version and publisher gets -5.\newline
Article Reportedwrong(Rejected) : -5 for the user who reported.\newline
Article/comment upvote/downvote: +/-2
\section{Glimpse of Implementation of the System}
The system has been implemented in a very straightforward manner.The implementation of upvote,downvote button and flag for reporting the article is merged with the existing template.
Usage of these buttons will change the number of upvotes and downvotes in the version of the article and then according to the condition specified in grading system, user reputation will change accordingly as that moment only.\newline
There are several case studies which will be based on different scenarios[8] : \newline \newline
1. \textbf{Person Joining the community} : The user gets +25 reputation on his dashboard and his system reputation also increases by +25 at this moment.If the user is new to the system , then he gets +25 as the starting reputation.\newline \newline
2. \textbf{Person editing the article} : The person reputation increases as soon as he hits "save" button which inturn increases his reputation by +2 for initial implementation.\newline \newline
3. \textbf{Person reporting the article} : The person who reported the article has to give the reason of reporting the article.This article goes in the tab of "Articles Reported" where users of that community can view that reported articles.The community admin will have an additional option of approving or rejecting the flag.\newline
\begin{itemize}
    \item If he approves the request, the reputation of publisher goes down by 5 and user of that last edited version by 5.The reputation of person who reported increases by 5.
    \item  If he rejects the request, the reputation of person who reported will be decreased by 5.
\end{itemize}
4. \textbf{Person changing to publishable state} : The publisher gets the notification to review the article.If publisher accepts the article and publishes the article, then the publisher would get +5 points,author +5.

\section{Why all contributors who helped in editing should get equal points? }
The system consists of various versions of article.Our basic aim to select those versions which actually improvised the article in a better fashion. 
Let us consider $x_{1}$,$x_{2}$....$x_{n}$ as the number of versions after the version of first draft as $x_{0}$.\newline
Consider  $u_{k}$ : upvote of the $k^{th}$ version
And $d_{k}$ : downvote of the $k^{th}$ version\newline
Let the selected versions be denoted by $s_{i}$. U() is upvote and D() is the downvote.\newline
If the factor of U($x_{i}$) - U($s_{i-1}$)/abs(D($ s_{i-1}$) -D($ x_{i} $)) is greater than 1 then this means that there is a substantial increase in upvotes as compared to increase or decrease in downvotes. So we select only those versions which have the positive value of this term.
\subsection{(Selection Algorithm) Algorithm for selecting the versions which led to the improvement of the article on the basis of votes}
Let S be the set denoting the selected version of article for point allocation including $x_{0}$. The last element of the S set is represented by $S(s_{k})$.\newline
\begin{algorithm}[]
\caption{Algorithm for selecting better version}
\SetAlgoLined
\hrule
\vspace{6pt}
 $i = j = k = 0$\\
 $S = S \cup x_{0}$\\
 \While{$i \leq n-1$}{
  \If {$U(x_{i}) > U(S(s_{k})) and \frac{U(x_{i}) - U(S(s_{k}))}{abs( D(S(s_{k})) - D(x_{i}) )} > 1$ }{$
   S = S \cup x_{i}$\\
   $k = k+1$\\}
   $i = i+1$}
   \hrule
\vspace{6pt}
\end{algorithm}
\subsection{Point Allocation for versions of article in S} 
Now the ratio of ($u_{j}$+$d_{j}$) : {number of views} is compared and if the ratio is close to one then it is a good article else the article is not good as the previous one because number of views are way higher than $u_{j}$+$d_{j}$ suggesting that it was less relevant.\newline
Obviously $u_{j}$+$d_{j}$ can never be less than number of views. If this is true then something is wrong in code which is trivial to understand.\newline
So according to the tested statistics the allocation is :
We maintain a bank where the number of versions selected as $s_{i}$ will each have +5 point so bank has 5n points if n are selected.\newline\newline
Case 1: So those articles which have ratio close to one will have 70\% of 5n.\newline
Case 2: Publisher will have 20 \% of 5n.\newline
Case 3: And rest goes to remaining versions of article.\newline

Now further if there are two or more articles in every case then equal distribution will be followed.
The proximity of close to 1 can be decided by the creator of the system but the above tested results are for proximity -0.5 to 0.5 precision.
\subsubsection{Proof of the above allocation system}
The proof describes that the factor of $U(x_{i}) - U(s_{i-1})/abs(D( s_{i-1}) -D( x_{i} ))$ is the improvement factor which will always ensure the selection of those articles which have higher peaks leading it to grow to the publishable state.\newline
If the $x_{0}$ has  $U(x_{0})$, $D(x_{0})$,then the next selection on the basis of the algorithm ensures that if the factor is greater than 1 then there is an increase in upvotes more than that of downvotes. So this ensures the improvement in the article and hence it gets selected.\newline
The proof will be done by using the principle of mathematical induction.\newline
Base: If the draft is $x_{0}$,the selection of next version is basically $U(x_{i})-U(x_{0})/abs(D(x_{0}) - D(x_{i}))$ is greater than 1 which implies that the users of the community have liked the version more than that of the draft itself.\newline
This ensures that if there is an increase in upvote is more than decrease or increase in downvote ,then we should select that version i. This selection proves the improvement factor selection algorithm.\newline
Induction : If the previous selected version is $s_{k}$ then,the selection of next version is basically $U(x_{i}) - U(s_{k})/abs(D(s_{k}) - D(x_{i})) > 1$.\newline
This ensures that more increase in upvote than change in downvote is there which indicates that we should select the version i if the above condition is satisfied thus making it close to the the publishable model. This selection proves the improvement factor selection algorithm.
\section{Way to help Publisher reviewing the publishable state}
The detection of the versions which helped the article to improve has one more application. Every time , when the user changes the state to publishable, then the publisher just needs to keep a track of only those versions which are selected using the algorithm of selection which in turn also suggests that if there are any requests at which no version is selected, then publisher should directly reject that request.
\section{Conclusion}
Thus, this paper addresses the solution to two problems : \\
Fair Allocation of points to contributors and Reducing publisher effort to check all the articles.\\
The major problem of fair allocation of points among the contributors was that we have to analyze the improvement in the versions of an article. So, this problem of allocation of points to the contributors is presented by an algorithm with a theoretical proof above.\newline
The publisher has to approve whether the content is ready to be published or not. Authenticity of such a public content becomes easy due to the selection algorithm which can handle the requests that user might have done by clicking it to publishable state. If the algorithm includes this version, it becomes to the publishable state else the request is rejected.  \newline

\section{References}
\noindent[1] Dana Movshovitz-Attias, Yair Movshovitz-Attias, Peter Steenkiste, Christos Faloutsos, "Analysis of the Reputation System and User Contributions on a Question Answering Website: StackOverflow", 2013 IEEE/ACM International Conference on Advances in Social Networks Analysis and Mining.\newline
[2] Ashton Anderson, Daniel Huttenlocher, Jon Kleinberg, Jure Leskovec, "Discovering Value from Community Activity on Focused Question Answering Sites: A Case Study of Stack Overflow",  Proceedings of the 18th ACM SIGKDD international conference on Knowledge discovery and data mining, 2012\newline
[3] Andrew Marder,  "Stack Overflow Badges and User Behavior: An Econometric Approach",  Proceedings of the 12th Working Conference on Mining Software, 2015\newline
[4] Bo Adler, Luca de Alfaro, Ashutosh Kulshreshtha, Ian Pye, "Reputation Systems for Open Collaboration",  Communications of the ACM 2010, Copyright 200\newline
[5] Page Ranking Algorithm, text is available under the Creative Commons Attribution-ShareAlike License,\\ https://en.wikipedia.org/wiki/PageRank\newline
[6] Josang, A., Ismail, R. (2002, June). The beta reputation system. In Proceedings of the 15th bled electronic commerce conference (Vol. 5, pp. 2502-2511).\newline
[7] A Josang, R Ismail - Proceedings of the 15th bled electronic "The beta reputation system", 2002 - domino.fov.uni-mb.si\newline
[8] M Gupta, P Judge, M Ammar - of the 13th international workshop, "A reputation system for peer-to-peer networks", 2003- dl.acm.org\newline
[9] C. Boyd A. Josang, R. Ismail. A survey of trust and
reputation systems for online service provision.
Decision Support Systems, 43:618-644, 2007\newline
[10] B.T. Adler, K. Chatterjee, L. de Alfaro, M. Faella,
I. Pye, and V. Raman. Assigning trust to Wikipedia
content. InProc. of WikiSym 08:  International
Symposium on Wikis. ACM Press, 2008
\end{document}